\documentclass[preprints,article,accept,moreauthors,pdftex]{Definitions/mdpi} 

\usepackage{amsmath,amssymb,cases,bm}
\firstpage{1} 
\pubvolume{1}
\issuenum{1}
\articlenumber{0}
\pubyear{2020}
\copyrightyear{2020}
\history{Received: date; Accepted: date; Published: date}

\pdfoutput=1

\Title{Self-evaporation dynamics of quantum droplets in a $^{41}$K-$^{87}$Rb mixture}

\Author{Chiara Fort $^{1,2}$\orcidA{} and Michele Modugno $^{3,4}$*\orcidB{}}

\AuthorNames{Chiara Fort and Michele Modugno}

\address{%
$^{1}$ \quad LENS and Dipartimento di Fisica e Astronomia, Universit\`{a} di Firenze, 50019 Sesto Fiorentino, Italy\\
$^{2}$ \quad Istituto Nazionale di Ottica, CNR-INO, 50019 Sesto Fiorentino, Italy\\
$^{3}$ \quad Department of Physics, University of the Basque Country UPV/EHU, 48080 Bilbao, Spain\\
$^{4}$ \quad IKERBASQUE, Basque Foundation for Science, 48009 Bilbao, Spain}

\corres{Correspondence: michele.modugno@ehu.eus}

\abstract{We theoretically investigate the self-evaporation dynamics of quantum droplets in a $^{41}$K-$^{87}$Rb mixture, in free-space. The dynamical formation of the droplet and the effects related to the presence of three-body losses are analyzed by means of numerical simulations. We identify a regime of parameters allowing for the observation of the droplet self-evaporation in a feasible experimental setup.}


\begin{document}

\section{Introduction}

As demonstrated in a seminal paper by D. Petrov \cite{petrov2015}, ultracold quantum gases can exist in the form of self-bound droplets that do not expand even in the absence of any confinement. This liquid-like behavior, which originates from the interplay of attractive mean-field interactions and the repulsive effect of quantum fluctuations \cite{petrov2018,ferrier-barbut2018,ferrier-barbut2019}, was successfully observed in a number of experiments with dipolar condensates \cite{ferrier-barbut2016,ferrier-barbut2016b,schmitt2016,chomaz2016,tanzi2019}, homonuclear mixtures of $^{39}$K \cite{cabrera2018,semeghini2018,cheiney2018,ferioli2018}, and recently in a heteronuclear mixture of $^{41}$K and $^{87}$Rb \cite{derrico2019,burchianti2019}. These findings have triggered an intense research activity on droplets properties (see, e.g., the recent reviews in Refs. \cite{bottcher2020,luo2020}) and on the so-called Lee-Huang-Yang fluid \cite{jorgensen2018,minardi2019,skov2020}, and have motivated further studies in low-dimensional systems \cite{morera2020,lavoine2020} and also beyond Petrov's theory \cite{hu2020a,hu2020b,hu2020c,zin2020,ota2020}.

Besides being self bound, one of the most peculiar characteristics of these quantum objects is the fact that they can be \textit{self-evaporating}. Namely, in certain regimes the droplet cannot sustain any collective mode, such that any initial excitation is completely dissipated (that is, \textit{evaporated}) until the system relaxes into a droplet with a lower number of atoms. This remarkable feature, originally predicted in Ref. \cite{petrov2015} for bosonic mixtures, has been so far elusive to the experimental detection. Indeed, the first pioneering experiments performed with bosonic mixtures \cite{cabrera2018,semeghini2018} have shown that the homonuclear mixtures of $^{39}$K suffer from strong three-body losses that continuously drive the system out of equilibrium, and eventually lead to the depletion of the droplet. This behavior was then confirmed by the theoretical analysis reported in Ref. \cite{ferioli2019}, where the complex dynamics taking place during the droplet formation is thoroughly described. There, it was shown that the evolution of the system is indeed dominated by the presence of three-body losses and by a continuous release of atoms to restore the proper population ratio, whereas the self-evaporation mechanism plays only a negligible role, if any.

In this respect, the recent experiment in Ref. \cite{derrico2019} represents a promising setup in which the self-evaporation mechanism could be properly investigated. As a matter of fact, for a $^{41}$K-$^{87}$Rb mixture the regime of parameters for which self-bound droplets form is such that three-body losses are expected to be significantly suppressed. Indeed, in such a mixture droplets form at lower densities $n\sim(\delta g/g)^2 a^{-3}$, and this allows for a larger ratio $\tau_\mathrm{life}/\tau \sim n^{-1}$, with $\tau_\mathrm{life}$ and $\tau$ being respectively the lifetime (limited by three-body losses) and the characteristic time scale of the droplet dynamics \cite{derrico2019}. Thus, the fact that the $^{41}$K-$^{87}$Rb mixture is characterized by larger scattering lengths with respect to the $^{39}$K mixture employed in Refs. \cite{cabrera2018,semeghini2018} allows for lower densities and, therefore, longer lifetimes. Actually, in that experiment no appreciable effect of three-body losses was observed on the timescale of several tens of milliseconds.

Motivated by the previous discussion, in this paper we analyze the self-evaporation mechanism and the role of three-body losses on the collective excitations of a $^{41}$K-$^{87}$Rb quantum droplet.
For the sake of conceptual clarity, and in view of the fact that quantum droplets at equilibrium are spherically symmetric objects, we shall restrict the analysis to the case of a spherically symmetric system. Though this assumption limits the study to the monopole collective mode, it is however sufficient to draw the general behavior of the system and to discuss the role of three-body losses. 

The paper is organized as follows. In Section \ref{sec:model} we review the general formalism for describing quantum droplets in heteronuclear bosonic mixtures. Then, in Section \ref{sec:preparation} we discuss the preparation of the initial state, namely a droplet compressed by a harmonic trapping. 
The dynamics of the droplet after the release of the trap is then studied in Section \ref{sec:dynamics} in the unitary case (no losses), and in the presence of three-body losses. Finally, in Section \ref{sec:conclusions} we draw the conclusions.

\section{Self-bound droplets}
\label{sec:model}

We consider a binary condensate of $^{41}$K and $^{87}$Rb atoms in the $\left|F=1,m_{F}=1\right\rangle$ state, as considered in the recent experiment in Ref. \cite{derrico2019}. The two components will be indicated as 1 and 2, respectively. 
This system is described by the following Gross-Pitaevskii (GP) energy functional, including both the mean field term and the Lee-Huang-Yang (LHY) correction accounting for quantum fluctuations in the local density approximation: \cite{ancilotto2018}
\begin{equation}
E = \sum_{i=1}^{2}\int \left[\frac{\hbar^2}{2m_i}|\nabla \psi_{i}(\bm{r})|^2  + V_{i}(\bm{r}) n_{i}(\bm{r})\right]d\bm{r}
+\frac{1}{2}\sum_{i,j=1}^{2}g_{ij}\int n_{i}(\bm{r})n_{j}(\bm{r})d\bm{r}
+\int {\cal E} _{\rm LHY}(n_1(\bm{r}),n_2(\bm{r}))d\bm{r}\,,
\label{eq:energyfun}
\end{equation}
where $m_{i}$ are the atomic masses,  $V_i(\bm{r})$ the external potentials, and $n_i(\bm{r})=|\psi _i(\bm{r})|^2$ the densities of the two components ($i=1,2$). The LHY correction reads \cite{petrov2015}
\begin{align}
{\cal E} _{\rm LHY} &= \frac{8}{15 \pi^2} \left(\frac{m_1}{\hbar^2}\right)^{3/2}
\!\!\!\!\!\!(g_{11}n_1)^{5/2} 
f\left(\frac{m_2}{m_1},\frac{g_{12}^2}{g_{11}g_{22}},\frac{g_{22}\,n_2}{g_{11}\,n_1}\right)
\nonumber
\\
&\equiv \kappa  (g_{11}n_1)^{5/2}f(z,u,x),
\end{align}
with  $\kappa=8 m_1^{3/2}/(15 \pi^2 \hbar^3)$ and $f(z,u,x)>0$ being a dimensionless function of the parameters $z\equiv m_2/m_1$, $u\equiv g_{12}^2/(g_{11}g_{22})$, and $x \equiv g_{22}n_2/(g_{11}n_1)$  \cite{petrov2015,ancilotto2018}.
The mixture is completely characterized in terms of the intraspecies $g_{ii}=4\pi \hbar^2 a_{i}/m_{i}$ ($i=1,2$), and interspecies $g_{12}=2\pi \hbar^2 a_{12}/m_{12}$ coupling constants,  where $m_{12}=m_{1}m_{2}/(m_{1}+m_{2})$ is the reduced mass. 
The values of the homonuclear scattering lengths are fixed to $a_{11}=62 a_0$ \footnote{A. Simoni, private communication.}, $a_{22}=100.4 a_0$, whereas the heteronuclear scattering length $a_{12}$ is considered here as a free parameter, that can be tuned by means of Feshbach resonances \cite{derrico2019}. The onset of the MF collapse regime corresponds to $\delta g = g_{12}+\sqrt{g_{1}g_{2}}=0$, at $a_\mathrm{12}^{c}=-73.6a_0$ \cite{riboli2002}. 

In free space ($V_{i}\equiv0$), the equilibrium density of a droplet is obtained by requiring the vanishing of the total pressure \cite{astrakharchik2018}, which yields \cite{petrov2015}
\begin{align}
n^{0}_{1}&=\frac{25\pi}{1024}\frac{1}{a_{11}^3}\frac{\delta g^2}{g_{11}g_{22}}
f^{-2}\left(\frac{m_2}{m_1},1,\sqrt{\frac{g_{22}}{g_{11}}}\right)
\\
n^{0}_{2}&=n^{0}_{1}\sqrt{\frac{g_{11}}{g_{22}}}.
\label{eq:density1}
\end{align}
Following \cite{petrov2015,ancilotto2018}, we consider this function at the mean-field collapse $u=1$, $f(z,1,x)$.
We note that the actual expression for $f$ can be fitted very accurately with the same functional form of the homonuclear case \cite{minardi2019}
\begin{equation}
f\left(\frac{m_2}{m_1},1,\sqrt{\frac{g_{22}}{g_{11}}}\right)\simeq \left[1+\left(\frac{m_2}{m_1}\right)^{3/5}\!\!\!\! \sqrt{\frac{g_{22}}{g_{11}}}\right]^{5/2}.
\end{equation}

For a finite number of
atoms the droplet has a finite size, and it can be effectively described by a single wave function that
satisfies the following dimensionless equation \cite{petrov2015}
\begin{equation}
\left[-\frac12\nabla_{\tilde{\bm{r}}}^{2} - 3\widetilde{N}|\phi_{0}|^{2} +\frac52\widetilde{N}^{3/2}|\phi_{0}|^{3}\right]\phi_{0}=\tilde{\mu}\phi_{0},
\label{eq:droplet}
\end{equation}
with $\int|\phi_{0}|^{2}d\tilde{\bm{r}}=1$, and 
\begin{equation}
N_{i}=n_{i}^{(0)}\xi^{3}\widetilde{N},
\label{eq:Ni}
\end{equation}
where $\sum_{i}N_{i}=N$ and $N_{1}/N_{2}=\sqrt{g_{22}/g_{11}}$ [see Eq. (\ref{eq:density1})].
Here $\xi$ represents the length scale of the droplet \cite{petrov2015,hu2020c}
\begin{align}
\xi&=\hbar
\left[
\frac{3}{2}\frac{\sqrt{g_{11}}/m_{2}+\sqrt{g_{22}}/m_{1}}{|\delta g|\sqrt{g_{11}}n_{1}^{(0)}}
\right]^{1/2},
\end{align}
and $n_{i}^{(0)}$ the equilibrium density of each component in the uniform case. We also remind that the wave functions of the two condensates forming the droplet are given by $\psi_{i}=\sqrt{n_{i}^{(0)}}\phi_{0}$.

\section{Preparation of the initial state}
\label{sec:preparation}

Following Ref. \cite{derrico2019}, we initially prepare the atomic mixture in the ground state of an optical dipole trap. 
In order to simplify the discussion, and motivated by the fact that quantum droplets at equilibrium are spherically symmetric objects, we assume i) that the two condensates are initially prepared in a spherically symmetric potential $U^{d}_{j}(r)$ ($j=1,2$), and ii) that the differential vertical gravitational sag (due to the different masses of the two atomic species) can be exactly compensated.
With these assumptions, which will be maintained throughout this work for easiness of calculations and conceptual clarity, both the ground state and the dynamical evolution can be obtained by solving spherically symmetric equations. We remark that this approach restricts the analysis to the excitations of the monopole mode only. Indeed, the excitation of surface modes (with angular momentum $\ell\neq0$), which arise when the condensate is prepared in an asymmetric trap, is ruled out by the assumption of spherical symmetry. However, the presence of these modes is not expected to modify qualitatively the picture resulting from the following discussion.

The ground state of the system is obtained by minimizing the energy functional in Eq. (\ref{eq:energyfun}) by means of a steepest-descent algorithm \cite{press2007}. Formally, this corresponds to solve the following set of stationary Gross-Pitaevskii (GP) equations for two wave functions $\psi_j$ \cite{dalfovo1999}
\begin{numcases}{}
\left[-\frac{\hbar^2}{2m_{1}}\nabla^{2}_{r} + U^{d}_{1}(r)
+ \mu_{1}(n_{1},n_{2}) \right]\psi_{1}=\mu_{1}\psi_{1}
\nonumber
\\
\label{eq:2gpe-static}
\\
\left[-\frac{\hbar^2}{2m_{2}}\nabla^{2}_{r} + U^{d}_{2}(r)
+ \mu_{2}(n_{1},n_{2}) \right]\psi_{2}=\mu_{2}\psi_{2},
\nonumber
\end{numcases}
where $\nabla_{r}^{2}$ represents the radial component of the Laplacian:
\begin{equation}
\nabla_{r}^{2}={\frac{1}{r^{2}}}{\frac{\partial }{\partial r}}\left( r^{2}{
\frac{\partial }{\partial r}}\right) ={\frac{\partial^{2}}{\partial r^{2}}}+
{\frac{2}{r}}{\frac{\partial }{\partial r}}.    
\end{equation}
The local chemical potentials $\mu_{i}(n_{1},n_{2})$ include both the usual mean-field term and the Lee-Huang-Yang (LHY) correction \cite{petrov2015}, namely 
\begin{equation}
\mu_{i}(n_{1},n_{2}) = g_{ii}n_{i} + g_{12}n_{j} + \frac{\delta E_{LHY}}{\delta n_{i}},\quad i\neq j,
\end{equation}
with $n_{i}(r)\equiv|\psi_{i}(r)|^{2}$. 

We recall that the optical dipole potential experienced by the two atomic species can be written as $U^{d}_{j}(r)= U_{0j}I(r)$, where $U_{0j}$ is species-dependent ($j=1,2$) and it depends on the atomic species polarizability and the wavelength of the optical potential, and $I(r)$ represents the intensity of the laser beam. For the atomic mixture $^{41}$K-$^{87}$Rb and a wavelength $\lambda=1064$ nm \cite{derrico2019} one has $\alpha\equiv U_{02}/U_{01}\simeq1.15$. Then, in the harmonic approximation, we have
\begin{equation}
U^{d}_{j}(r)=\frac12m_{j}\omega^{2}_{j}r^{2},
\end{equation}
with $\omega_{1}=\omega_{2}\sqrt{m_2/(\alpha m_1)}$. In the following, the value of the trap frequency will be considered as a free parameter, to be tuned in order to produce the desired amount of initial excitations. For low enough $\omega_j$, the droplet will be prepared close to the equilibrium configuration in free space, whereas a tight confinement obviously corresponds to the excitation of a compressional mode.

The other free parameters of the system are the inter-species scattering length $a_{12}$, that we will consider in the range $a_{12}\in[-95,-73.6]a_0$, below to the mean-field collapse threshold, and the number of atoms of the two species. As for the latter, we shall vary $N_{2}$ in the range $[1.5,15]\times10^4$, keeping $N_{1}=N_{2}\sqrt{g_{22}/g_{11}}$, so that the atom numbers match the nominal equilibrium ratio in Eq. (\ref{eq:density1}).

\section{Dynamics}
\label{sec:dynamics}

Once the initial state is prepared, the optical dipole trap is switched off and the system is let to evolve.
The droplet dynamics is studied here by solving the following set of GP equations \cite{wachtler2016a,wachtler2016b,ferioli2019}
\begin{numcases}{}
i\hbar\partial_{t}\psi_{1}= \left[-\frac{\hbar^2\nabla^{2}_{r}}{2m_{1}} +\mu_{1}(n_{1},n_{2}) -i \hbar \frac{K_{3}}{2} |\psi_{2}|^{4}\right]\psi_{1}
\nonumber
\\
\label{eq:2gpe+k3}
\\
i\hbar\partial_{t}\psi_{2}= \left[-\frac{\hbar^2\nabla^{2}_{r}}{2m_{2}} +\mu_{2}(n_{1},n_{2}) -i \hbar K_{3} |\psi_{1}|^{2}|\psi_{2}|^{2}\right]\psi_{2},
\nonumber
\end{numcases}
in the presence of three-body losses, which are included by adding to the energy functional in Eq.~(\ref{eq:energyfun}) a dissipative term
$-(i/2)\hbar K_3 \int n_1(\bm{r},t)n_2(\bm{r},t)^{2} d^{3}r $
(with $K_3=7\times 10^{-41}$ m$^6$/s, see Ref. \cite{derrico2019})
accounting for the dominant recombination channel, \textit{i.e.} K-Rb-Rb.

\begin{figure}[t]
\centerline{\includegraphics[width=0.6\columnwidth]{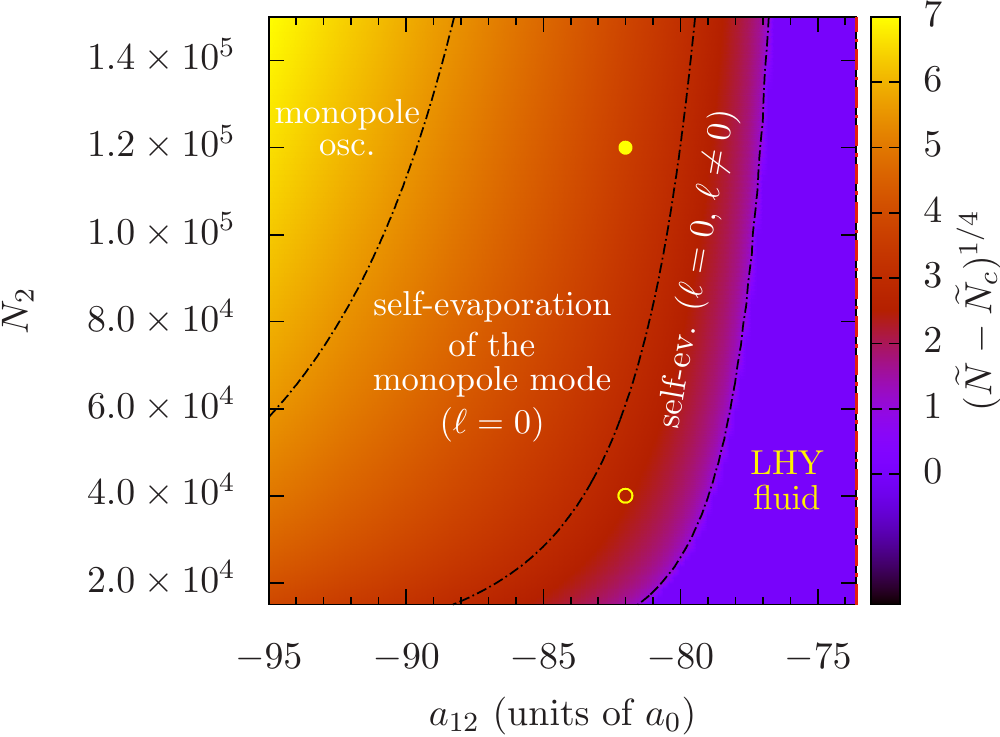}}
\caption{Heat map of $(\widetilde{N}-\widetilde{N}_{c})^{1/4}$ as a function of $a_{12}$ and $N_2$, with $\widetilde{N}_{c}=18.65$ being the critical value for the existence of a droplet \cite{petrov2015}. The dotted-dashed lines, corresponding to $\widetilde{N}=933.7$, $94.2$, $18.65$ (from left to right), 
represent the boundaries between the different regimes (see text). The vertical red dashed-dotted line at $a_{12}=-73.6$ corresponds to the onset of the MF collapse.
The two circles at $a_{12}=-82a_{0}$ refer to the parameter configurations considered in Sec. \ref{sec:dynamics}.}
\label{fig:ntilde}
\end{figure}

Since the droplet is prepared in a \textit{compressed} configuration (owing to the presence of the trap), the droplet will start oscillating.
According to discussion in Ref. \cite{petrov2015}, in the absence of three-body losses the dynamics is expected to be characterized either by sinusoidal oscillations of the droplet width, where the monopole mode exists, or by damped oscillations, in the so-called \textit{self-evaporation} regime. The latter represents one of the remarkable properties of quantum droplets \cite{petrov2015,ferioli2019}, and it  takes place in a certain window of $\widetilde{N}$, where the excitation spectrum of the droplet lies entirely in the continuum. The nominal phase diagram for our system (obtained from the predictions Ref. \cite{petrov2015}) is shown in Fig. \ref{fig:ntilde} as a function of the interspecies scattering length $a_{12}$ and of the (initial) number of atoms $N_2$. The specific combination of parameters used in the figure, $(\widetilde{N}-N_{c})^{1/4}$, is introduced after Ref. \cite{petrov2015}.
The monopole mode is expected to be stable for $\widetilde{N}>933.7$, and to evaporate for lower values of $\widetilde{N}$. In the window $94.2<\widetilde{N}<933.7$ other modes with $\ell\neq0$ (not included in the present discussion) may appear, whereas below $\widetilde{N}<94.2$ neither the monopole nor the surface modes ($\ell\neq0$) can be sustained, such that the droplet is expected to evaporate any initial excitation \cite{petrov2015}. Below $\widetilde{N}_{c}$, where a droplet cannot be formed, the mixture forms a so-called LHY fluid: in this regime the MF interactions almost cancel out ($\delta g/\sqrt{g_{11}g_{22}}\approx0$), and the system is governed only by quantum fluctuations \cite{jorgensen2018,minardi2019,skov2020}.
%

\subsection{Estimate of the droplet lifetime}
\label{sec:3bl}

Before  analyzing the detailed dynamical behavior of our system it is convenient to discuss the role of three-body losses on the droplet lifetime. To this aim, we shall consider a droplet in free space, at equilibrium (at $t=0$). Then, \textit{assuming} that the shape of the droplet is weakly affected by the atom losses (which is justified at the early stages at the evolution, at least),  we can write 
\begin{equation}
n_i(r,t)\simeq N_i(t)\rho(r),   
\end{equation}
where $N_{1}(0)/N_{2}(0)=\sqrt{g_{22}/g_{11}}\equiv\gamma$ , and with $\rho(r)$ representing the droplet density profile normalized to unity [see Eqs. (\ref{eq:density1}) and (\ref{eq:Ni})].
With this in mind, the evolution of $N_i(t)$ can be obtained from the previous Eqs. (\ref{eq:2gpe+k3}) by neglecting the kinetic and chemical potential terms [which concur in determining the shape $\rho(r)$], left multiplying each equation by $\psi_i^*$, 
and integrating over the volume (see also Ref. \cite{altin2011}), yielding
\begin{numcases}{}
\label{eq:eqlossN1}
\frac{d\bar{N}_{1}}{d\tau}= -\bar{N}_{1}\bar{N}_{2}^{2}
\\
\frac{d\bar{N}_{2}}{d\tau}= -2\gamma \bar{N}_{1}\bar{N}_{2}^{2},
\label{eq:eqlossN2}
\end{numcases}
where we have defined $\bar{N}_i(\tau)\equiv N_i(\tau)/N_i(0)$ and $\tau\equiv t K_{3}N_2^2(0)\int\rho^{3}d^{3}r$. Notice that the factor of $2$ in Eq. (\ref{eq:eqlossN2}) corresponds to
the fact that here we are considering the dominant recombination channel K-Rb-Rb (two atoms of Rb are lost for each atom of K).
\begin{figure}[t]
\centerline{\includegraphics[width=0.8\columnwidth]{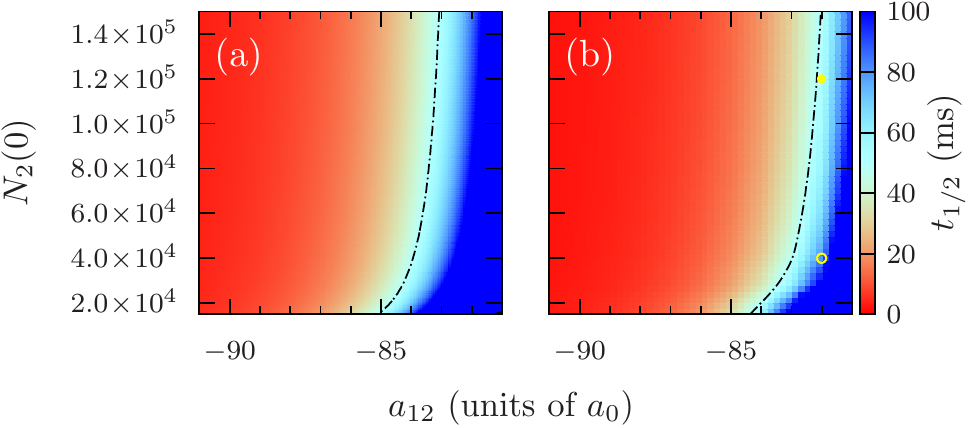}}
\caption{Heat map of $t_{1/2}$ as a function of $a_{12}$ and $N_2(0)$. (a) Prediction of the analytical model, see Eq. (\ref{eq:lossN2}); (b) values extracted from the actual decay of $N_2(t)$ obtained from the solution of the GP equations in (\ref{eq:2gpe+k3}). The color scale is saturated at $t_{1/2}=100$ ms (dark blue). The dashed-dotted line corresponds to $t_{1/2}=50$ ms. The two circles at $a_{12}=-82a_{0}$ refer to the parameter configurations considered in the GP simulations of Sec.~\ref{sec:monopole}.}
\label{fig:halflife}
\end{figure}
The above equations have an approximate solution of the form
\begin{align}
\label{eq:lossN1}
\bar{N}_1(\tau) &= \frac{1-\beta}{1+ (\tau/\tau_1)^{\alpha_1}} + \beta, 
\\
\bar{N}_2(\tau) &= \frac{1}{1+ (\tau/\tau_2)^{\alpha_2}},
\label{eq:lossN2}
\end{align}
which is very accurate, indeed.
In the present case ($\gamma\simeq0.8736$), from a fit of the exact numerical solution of Eqs. (\ref{eq:eqlossN1}) and (\ref{eq:eqlossN2}) we find $\tau_1\simeq0.70$, $\alpha_1\simeq0.88$,  $\beta\simeq0.43$, $\tau_2\simeq0.71$, $\alpha_2\simeq0.86$. In particular, $\tau_{2}$ corresponds to the \textit{half-life} of the $i=2$ component (regardless of the value of $\alpha_2$), in which losses are dominant. Therefore, it represents a characteristic time through which we can measure the impact of three-body losses on the lifetime of the droplet. In our simple model the half-life is therefore $t_{1/2}\simeq\tau_{2}/\left[K_{3}N_2^2(0)\int\rho^{3}d^{3}r\right]$ that, besides the explicit dependence on $N_2(0)$, also depends implicitly on $a_{12}$ through the density distribution $\rho(r)$. The behavior of $t_{1/2}$ as a function of  $a_{12}$ and $N_2(0)$ is shown in Fig. \ref{fig:halflife}. There we compare the prediction of the above analytical model with the actual values obtained from the solution of the GP equations in (\ref{eq:2gpe+k3}). The qualitative agreement is remarkable.

\subsection{Damped monopole oscillations}
\label{sec:monopole}

Given the above picture, in the following we shall investigate the self evaporation dynamics for $a_{12}=-82a_{0}$, where the droplet half-life is larger than $50$ ms [see Fig. \ref{fig:halflife}(b)]. In particular, we consider two different configurations with $N_{2}=4\times10^4$ and $N_{2}=1.2\times10^5$, indicated by the yellow circles in Figs. \ref{fig:ntilde} and \ref{fig:halflife}(b), both lying in the self-evaporation regime.
It is worth to remark that in the regime where the monopole mode is expected to be stable (see Fig. \ref{fig:ntilde}) the droplet is affected by severe three-body losses (see Fig. \ref{fig:halflife}), which make the detection of this mode unfeasible.

In order to distinguish between the atoms remaining in the droplet and those that evaporate, we define the \textit{droplet volume} as that contained within a certain bulk radius. In the present case, it can be conveniently fixed to $R_{d}=8$ $\mu$m \footnote{The numerical simulations of the GP equations are performed on a computational box that is at least one order of magnitude larger than $R_{d}$.}. Accordingly, with $N_{i}^{d}(t)$ we indicate the number of atoms of each species ($i=1,2$) within the droplet volume, at time $t$. Then, we define a \textit{running} value of $\widetilde{N}$ as
\cite{ferioli2019}
\begin{equation}
\widetilde{N}_{R}(t)
\equiv
\frac{1}{\xi^3}\frac{N_{1}^{d}(t)+N_{2}^{d}(t)}{N_{1} + N_{2}}
=k\sum_{i=1}^{2}N_{i}^{d}(t),
\end{equation}
where $k\simeq4.41\times10^{-3}$ for the current values of the scattering lengths.

\begin{figure}[t]
\centerline{\includegraphics[width=0.45\columnwidth]{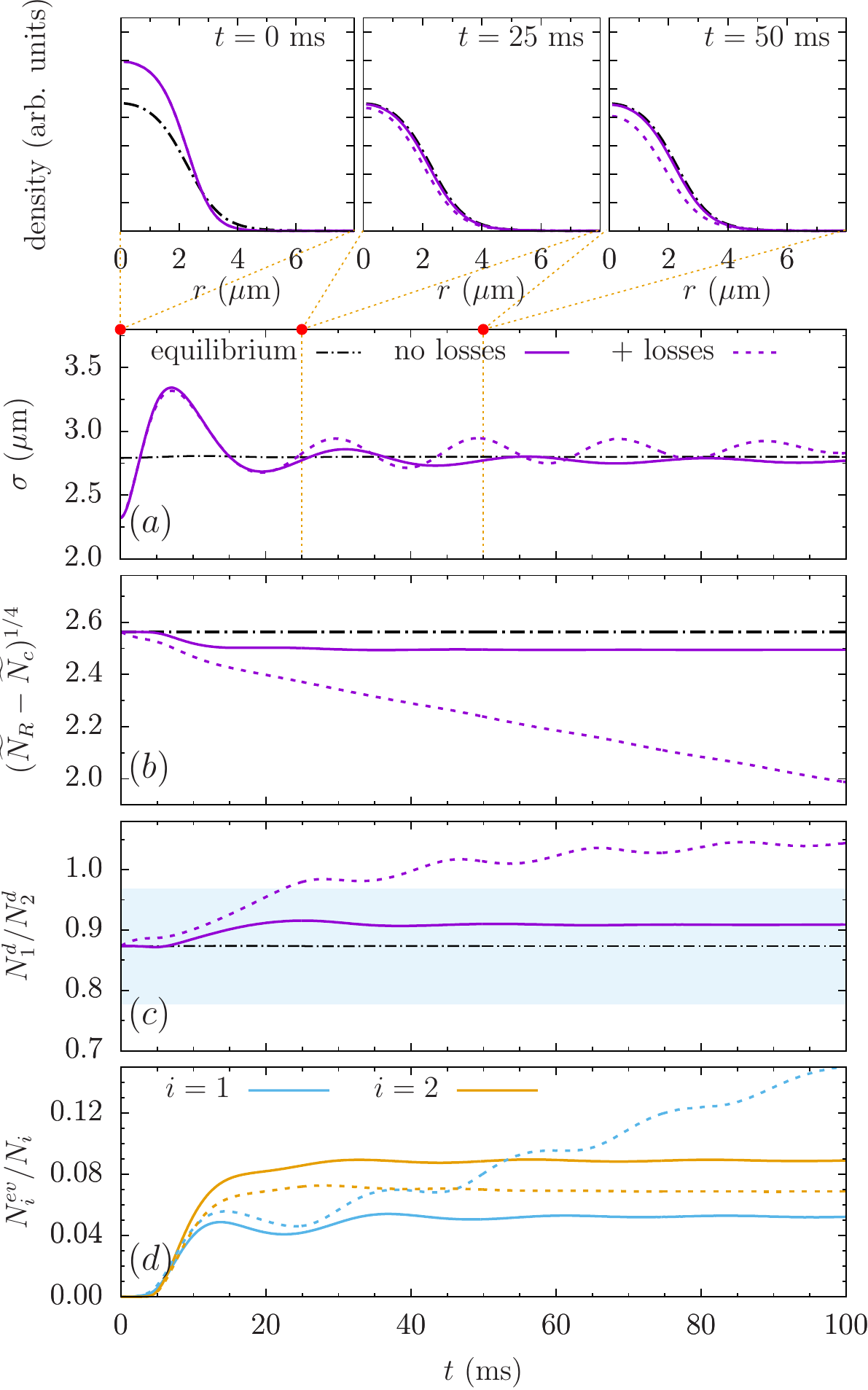}
\includegraphics[width=0.45\columnwidth]{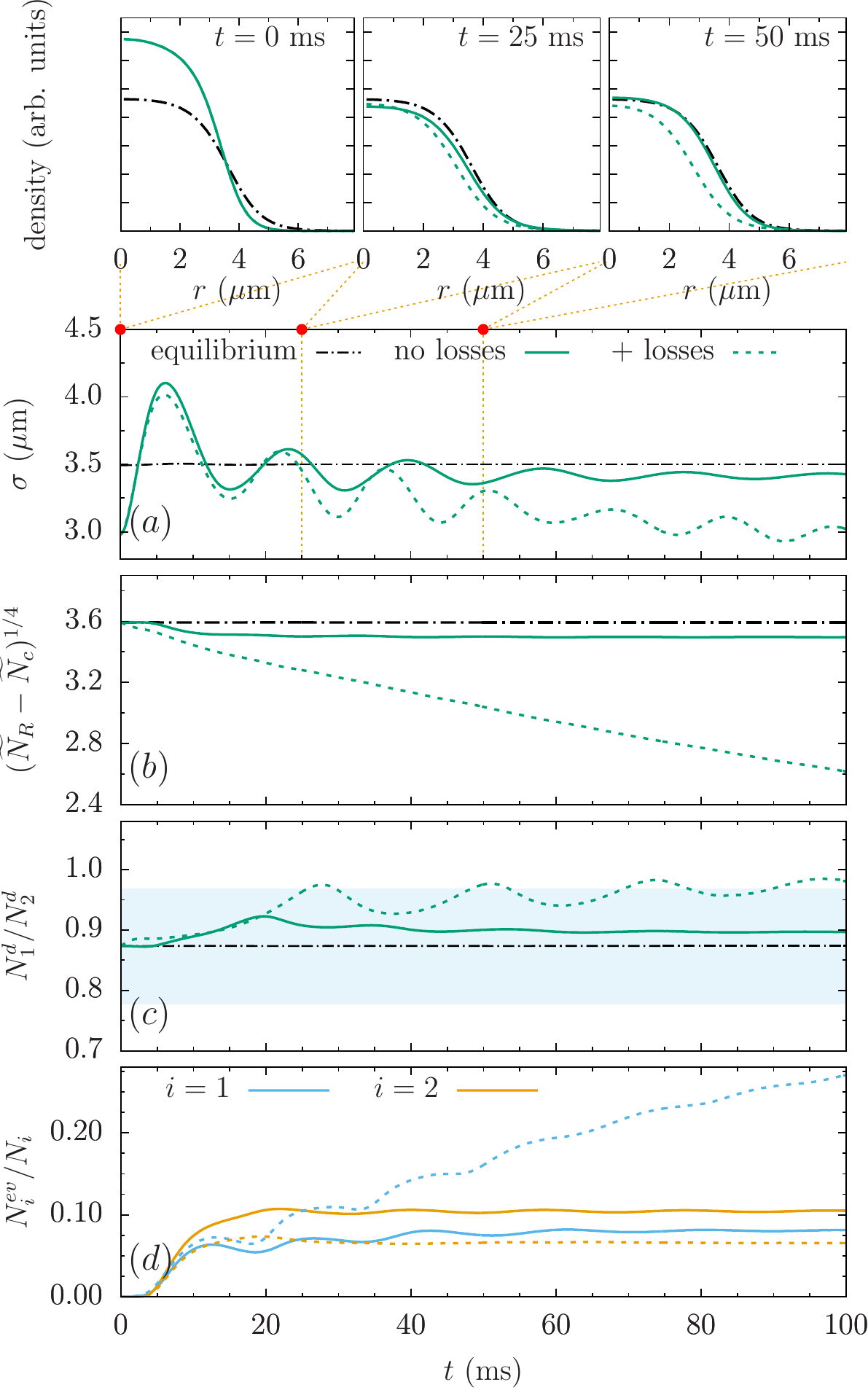}}
\caption{Evolution of (a) the droplet width $\sigma(t)$, (b) the running value of $\widetilde{N}_{R}(t)$ (see text), (c) the ratio $N_{1}^{d}(t)/N_{2}^{d}(t)$, and (d) the fraction of evaporated atoms for each species $N_{i}^{ev}(t)/N_{i}$, after the release from a trap of frequency $\omega_2=2 \pi \times 50$ Hz, with and without three-body losses. The insets in the top row show the total density of the binary mixture, $n(r,t)=\sum_{i=1}^{2}N_{i}|\psi_{i}(r,t)|^{2}$, at different evolution times, corresponding to the red circles in (a). 
The horizontal line in (b) represent the \textit{nominal} equilibrium value $N_1/N_2=\sqrt{g_{22}/g_{11}}\simeq0.873$, and the dashed area the corresponding tolerance (see text). Here  $a_{12}=-82a_{0}$, $N_{2}=4\times10^{4}$ (left), $N_{2}=1.5\times10^{5}$ (right).}
\label{fig3}
\end{figure}
The evolution of the system is shown in in Fig.~\ref{fig3}, where we plot the rms width $\sigma(t)$ of the droplet, the running value of $[\widetilde{N}_{R}(t)-N_{c}]^{1/4}$,  the ratio $N_{1}^{d}(t)/N_{2}^{d}(t)$, and the fraction of evaporated atoms for each species $N_{i}^{ev}(t)/N_{i}$, with and without three-body losses. In both cases the mixture is initially prepared in the ground state of a dipole trap of frequency $\omega_2=2 \pi \times 50$ Hz (we recall that $\omega_{1}=\omega_{2}\sqrt{m_2/(\alpha m_1)}$, see Sec. \ref{sec:preparation}; in the present case $\omega_1=2 \pi \times 67.9$ Hz).

Let us first focus on the clean case, in the absence of three-body losses. 
As expected, we find that in both the investigated cases
the droplet width performs damped sinusoidal oscillations (corresponding to a damped monopole mode), and the system eventually relaxes to an equilibrium configuration [see Fig. \ref{fig3}(a), along with the top panels], corresponding to a smaller, stationary value of $\widetilde{N}_R(t)$, see panels (b). This decrease in the number of particles is a consequence of the self-evaporation mechanism \cite{petrov2015,ferioli2019}, which
takes place within the first $15$ ms. Looking at panel (d), where we plot the fraction of atoms of each species that are lost by self-evaporation, it is clear that the values of $N_{i}^{ev}(t)/N_{i}$ soon reach their asymptotic value (modulo small fluctuations). 
In addition, also the ratio between the atom numbers in the two components remains close to the equilibrium value, see panels (c), with deviations below the reequilibration threshold (shaded area in the figure). Indeed, we recall that a droplet can sustain an excess of particles in one of the two components $\delta N_{i}^d/N_{i}^d$ up to a critical value $\sim \delta g/\sqrt{g_{11}g_{22}}$ \cite{petrov2015} ($\approx11\%$ in the present case), beyond which particles in excess are expelled. 

Let us now discus how the presence of three-body losses affects the above picture. From panels (b) it is evident that losses produce a continuous drain of particles from the droplet. Nevertheless, after $100$ ms of evolution the value of $\widetilde{N}_R(t)$ is still well above the critical value $\widetilde{N}_c$ for the existence of a droplet. 
Indeed, after several tens of milliseconds the mixture still forms a self-bound droplet (see e.g. the density profile at $50$ ms) which keeps undergoing damped monopole oscillations, as shown in panels (a). Notably, the initial stage of the evolution is still dominated by self-evaporation, see panel (d), and then both the frequency and the amplitude of the oscillations become larger than in the clean case (without losses),  because of the atoms that leave the bulk and populate the tails. 
Actually, there are two opposite effects taking place: on the one hand the droplet shrinks because of the loss of atoms due to three-body recombination, on the other hand there is an outward flow of K atoms that evaporate from the droplet, see panel (d). This species selective evaporation is due to the fact that the major loss of atoms affects the Rb component (see previous section), so that a progressively increasing fraction of K atoms cannot stay bound inside the droplet, and it is let free to expand. We remark that this is obviously a non-equilibrium process, as it is evident from the fact that $\widetilde{N}_R(t)$ does not relaxes to a stationary value and that the ratio $N_{1}^{d}(t)/N_{2}^{d}(t)$ soon run over the reequilibration threshold, anyway. All this is responsible for the different 'asymptotic' behavior of $\sigma(t)$ in the two cases shown in panels (a). For $N_{2}=4\times10^{4}$ (left), the two effects approximately compensate each other, and the width oscillates close to the nominal equilibrium value without losses. Instead, for $N_{2}=1.5\times10^{5}$ (right), which is characterized by a shorter lifetime, the droplet width decreases progressively (though keeping oscillating). It is worth to remark that the actual behavior is also sensitive to the choice of the droplet radius $R_d$.

\section{Conclusions}
\label{sec:conclusions}

We have theoretically investigated the self-evaporation dynamics of quantum droplets in a $^{41}$K-$^{87}$Rb mixture in feasible experimental setups, including the effects of three-body losses. The mixture is prepared in the ground state of a spherically symmetric harmonic trap, that is then released thus letting the system evolve in free space. The subsequent dynamics, characterized by the excitation of the monopole breathing mode, has been analyzed by solving the coupled Gross-Pitaevskii equations for the two components. For the estimated values of three-body losses ($K_3=7\times 10^{-41}$ m$^6$/s for the dominant recombination channel K-Rb-Rb \cite{derrico2019}), we find that by tuning the interspecies scattering length $a_{12}$, the lifetime of the system can be easily adjusted to be of the order of, or larger than, $100$ ms. This makes 
$^{41}$K-$^{87}$Rb droplets much more robust than those realized with two hyperfine states of $^{39}$K, whose lifetime is limited to the order of $10$ ms \cite{cabrera2018,semeghini2018,ferioli2019}. Such long lifetimes permits to follow the droplet dynamics for several tens of milliseconds, without any appreciable loss of resolution. 
In this scenario, we have found that the initial stage of the evolution is dominated by the self-evaporation mechanism even in the presence of three-body losses, and that the latter induce an interesting non equilibrium dynamics at later times. These findings make the experimental investigation of collective modes of self-bound droplets in $^{41}$K-$^{87}$Rb mixtures very promising.

\funding{This work was supported by the Spanish Ministry of Science, Innovation and Universities and the European Regional Development Fund FEDER through Grant No. PGC2018-101355-B-I00 (MCIU/AEI/FEDER, UE), by the Basque Government through Grant No. IT986-16.}

\acknowledgments{We thank Alessia Burchianti, Luca Cavicchioli, and Francesco Minardi for the critical reading of the manuscript.}

\begin{appendix}

\section{Numerical methods}

Let us consider a GP equation of the form 
\begin{equation}
i\partial_{\tau}\varphi(r,\tau) = H\varphi(r,\tau),
\end{equation}
with 
\begin{equation}
H = -\frac12\nabla_{r}^{2} + V(r) + g|\varphi(r,\tau)|^{2},
\end{equation}
and 
\begin{equation}
\nabla^2 = {\frac{\partial^2}{\partial r^2}} + \frac{2}{r}{\frac{%
\partial}{\partial r}}.
\end{equation}

The Crank-Nicholson algorithm (on a discrete space-time grid) consists in
solving the following system of linear equations with the components of the
vector $\varphi^{n+1}\equiv(\varphi_{1},\dots,\varphi_{j},\dots\varphi_{N})^{n+1}$ 
being the unknown variables \cite{press2007} 
\begin{equation}
\left( 1 + {\frac{i\Delta \tau}{2}} H\right)\varphi^{n+1}= 
\left( 1 - {\frac{i\Delta \tau}{2}} H\right)\varphi^{n}.  \label{eq:cn}
\end{equation}
Notice that this algorithm preserves the unitarity for real-time evolutions.

Derivatives can be written in terms of central differences \cite%
{abramowitz1964}, 
\begin{align}
\frac{\partial^2\varphi}{\partial r^2}& \to \frac{%
\varphi_{j+1}- 2\varphi_{j} + \varphi_{j-1}}{\Delta^2} \\
\frac{2}{r}\frac{\partial\varphi}{\partial r}& \to \frac{2}{r_{j}}\frac{%
\varphi_{j+1}- \varphi_{j-1}}{2\Delta}
\end{align}
that allow to write the linear system Eq. (\ref{eq:cn}) in the following
tridiagonal form, 
\begin{equation}
a_{j}\varphi^{n+1}_{j-1}+b_{j}\varphi^{n+1}_{j}+ c_{j}\varphi^{n+1}_{j+1} =
d_{j}.
\end{equation}

The explicit expression for the coefficients $a_{j}$, $b_{j}$, $c_{j}$ , and 
$d_{j}$ ($j=1,\dots N$) are 
\begin{align}
a_{j} &= {\frac{i\Delta\tau}{2}}\left[\frac{1}{\Delta^2}-\frac{1}{r_j\Delta}\right], \\
b_{j} &= 1 + {\frac{i\Delta\tau}{2}} \left[-\frac{2}{\Delta^2}+
V(r_{j}) + g|\varphi_{j}^{n}|^{2}\right], \\
c_{j} &= {\frac{i\Delta\tau}{2}}\left[\frac{1}{\Delta^2}+\frac{1}{r_j\Delta}\right], \\
d_{j} &= -a_{j}\varphi^{n}_{j-1}+(2-b_{j})\varphi^{n}_{j} -
c_{j}\varphi^{n}_{j+1},
\end{align}
which have to be accompanied by suitable boundary conditions in order to preserve the tridiagonal form.

\end{appendix}

\reftitle{References}
\externalbibliography{yes}
\bibliography{droplet}

\end{document}